\input harvmac

\lref\steve{S. Shenker, talk at Cargese workshop
on Random surfaces, Quantum Gravity, and Strings,
France (1990).}
\lref\joe{J. Polchinski, ``Dirichlet Branes and Ramond-Ramond
Charges'', {\it Phys. Rev. Lett.} {\bf 75} (1995) 4724,
hep-th/9510017.}
\lref\wsstinst{references on ws-st instanton dual examples}
\lref\joepart{J. Polchinski, ``Evaluation of the One Loop
String Path Integral'', {\it Comm. Math. Phys.} {\bf 104}
(1986) 37.}
\lref\eddyn{E. Witten, ``String Theory Dynamics in Various
Dimensions'', {\it Nucl. Phys.} {\bf B}443 (1995) 85.}
\lref\dkl{L. Dixon, V. Kaplunovsky, and J. Louis, 
``Moduli Dependence of String Loop Corrections to
Gauge Coupling Constants'',{\it Nucl. Phys.} {\bf B}355 (1991) 649.}
\lref\vw{C. Vafa and E. Witten, ``Dual String Pairs With
$N=1$ and $N=2$ Supersymmetry in Four Dimensions'', 
hep-th/9507050.}
\lref\gaugecond{S. Kachru and E. Silverstein, ``N=1 Dual
String Pairs and Gaugino Condensation'', {\it Nucl. Phys.}
{\bf B}463 (1996) 369, hep-th/9511228.}
\lref\dist{S. Shenker, ``Another Length Scale in String
Theory?'' }
\lref\banksdine{T. Banks and M. Dine, ``Coping with Strongly
Coupled String Theory'', {\it Phys. Rev.} {\bf D}50 (1994) 7454,
hep-th/9406132.}
\lref\gaillard{P. Binetruy, M.K. Gaillard, and Y. Wu, 
``Modular Invariant Formulation of Multi-Gaugino
and Matter Condensation'', hep-th/9611149.}
\lref\fhsv{S. Ferrara, J. Harvey, A. Strominger, and C. Vafa,
``Second Quantized Mirror Symmetry'', {\it Phys. Lett.}
{\bf B}361 (1995) 59. hep-th/9505162}
\lref\kv{S. Kachru and C. Vafa, ``Exact Results for
N=2 Compactifications of Heterotic Strings'',
{\it Nucl. Phys.} {\bf B}450 (1995) 69. hep-th/9505105.}
\lref\bbs{K. Becker, M. Becker, and A. Strominger, 
``Fivebranes, Membranes, and Nonperturbative String
Theory'', {\it Nucl. Phys.} {\bf B}456 (1995) 130, 
hep-th/9507158.}
\lref\sminst{E. Witten, ``Small Instantons in String Theory'',
{\it Nucl. Phys.} {\bf B}460 (1996) 541, hep-th/9511030.}
\lref\cdgp{P. Candelas, X. de la Ossa, P. Green, and L. Parkes,
``A Pair of Calabi-Yau Manifolds as an Exactly Soluble
Superconformal Theory'',{\it Nucl. Phys.} {\bf B}359 (1991) 21.}
\lref\mcclain{B. McClain and B. D. B. Roth, ``Modular
Invariance for Interaction Bosonic Strings at Finite Temperature'',
{\it Comm. Math. Phys.} {\bf 111} (1987) 539.}
\lref\harvmoore{J. Harvey and G. Moore, ``Five-brane
Instantons and $R^2$ Couplings in N=4 String Theory'',
hep-th/9610237.}
\lref\edsup{E. Witten, ``Non-perturbative Superpotentials
in String Theory''{\it Nucl. Phys.} {\bf B}474 (1996) 343,
hep-th/9604030.}
\lref\kss{S. Kachru, N. Seiberg, and E. Silverstein, ``SUSY
Gauge Dynamics and Singularities of $4d$ $N=1$ String Vacua'',
hep-th/9605036.}
\lref\ks{S. Kachru and E. Silverstein, ``Singularities, Gauge Dynamics, 
and Non-perturbative Superpotentials in String Theory'',
hep-th/9608194.}
\lref\specgeom{S. Ferrara and A. Strominger, in {\it Strings `89},
eds. R. Arnowitt, R. Bryan, M. J. Duff, D. V. Nanopoulos, and
C. N. Pope (World Scientific, 1989) 245;
A. Strominger, {\it Comm. Math. Phys.} {\bf 133} (1990) 163;
L.J. Dixon, V.S. Kaplunovsky, and J. Louis, {\it Nucl. Phys.} 
{\bf B}329 (1990) 27;
P. Candelas and X. de la Ossa, {\it Nucl. Phys.} {\bf B}355
(1991) 455; L. Castellani, R. D'Auria and S. Ferrara, 
{\it Phys. Lett.} {\bf B}241 (1990) 57; 
R. D'Auria, S. Ferrara, and P. Fre, 
{\it Nucl. Phys.} {\bf B}359 (1991) 705.}
\lref\klm{A. Klemm, W. Lerche, and P. Mayr, ``K3 Fibrations
and Heterotic Type II String Duality'', {\it Phys. Lett.}
{\bf B}357 (1995) 313.}
\lref\edsupII{R. Donagi, A. Grassi, and E. Witten, 
``A Nonperturbative Superpotential with $E_8$ Symmetry'',
{\it Mod. Phys. Lett.} {\bf A}11 (1996) 2199, 
hep-th/9607091.}
\lref\mayr{P. Mayr, ``Mirror Symmetry, $N=1$ Superpotentials,
and Tensionless Strings on Calabi-Yau Fourfolds'',
hep-th/9610162.}
\lref\katzvafa{S. Katz and C. Vafa, ``Geometric Engineering
of N=1 Quantum Field Theories'', hep-th/9611090.}
\lref\asplouis{P. Aspinwall and J. Louis, ``On the Ubiquity of
K3 Fibrations in String Duality'', {\it Phys. Lett}
{\bf B}369 (1996) 233.}
\lref\joeI{J. Polchinski, ``The Combinatorics of Boundaries
in String Theory'', {\it Phys. Rev.} {\bf D}50 (1994) 6041.}

\Title{RU-96-104, hep-th/9611195}
{\vbox{\centerline{Duality, Compactification, and $e^{-1/\lambda}$
Effects}
        \vskip4pt\centerline{in the Heterotic String Theory}}}
\centerline{Eva Silverstein 
\footnote{$^\star$}
{evas@physics.rutgers.edu 
}}
\bigskip\centerline{Department of Physics and Astronomy}
\centerline{Rutgers University}
\centerline{Piscataway, NJ 08855-0849}

\vskip .3in

Two classes of stringy instanton effects, stronger than
standard field theory instantons, are identified
in the heterotic string theory.  These contributions
are established using type IIA/heterotic and
type I/heterotic dualities.  They provide
examples for the heterotic case 
of the effects predicted by Shenker based on
the large-order behavior of perturbation theory.  The corrections
vanish as the radius of the compactification
goes to infinity.  For appropriate amplitudes, they are
computable worldsheet or worldline instanton effects 
on the dual side.  Some potential applications
are discussed.

\Date{11/96} 

\newsec{Introduction}

Nonperturbative effects in string theory
have come under significantly better control in recent years.
This progress has gone hand in hand with new understanding
of the degrees of freedom in the theory.  In particular,
D-branes realize concretely the 
``stringy instantons'' predicted by Shenker \steve\
in an expansion about weakly coupled type II and
type I string theories \refs{\joeI,\eddyn,\joe,\bbs}.  Shenker
argued that all closed string theories
should exhibit non-perturbative effects which fall
off like $e^{-1/\lambda}$ at weak coupling $\lambda$ in addition
to those understood from the low-energy effective field
theory which fall off like $e^{-1/\lambda^2}$.  Open
string theories automatically have such behavior. 
Type II string theories turn out to contain open
strings because of the presence of D-branes in the
nonperturbative spectrum, and their tension indeed
goes like $1/\lambda_{II}$.

There is one closed string theory left, for which
the argument in \steve\ applies but D-branes do
not appear: the heterotic string theory.  
From \steve\ we would expect $e^{-1/\lambda_{het}}$ effects even
in the highly symmetric decompactified ten-dimensional
theory.  In this note, we will find effects which
fall off like $e^{-f(M)/\lambda_{het}}$, where $f(M)$ is
a function of moduli which diverges as the size of
the compact space goes to infinity.

We find these effects in two different contexts,
making use of heterotic/Type IIA dualities and
heterotic/type I duality.  In each case the basic idea is
very simple.  In the four-dimensional
heterotic/type IIA dual pairs, the $4d$ heterotic
coupling $\lambda_{h,4}^2$ maps to $\alpha^\prime/A_{IIA}$, where $A_{IIA}$
is the area of a nontrivial 2-cycle C in the
compactification on the IIA side.  In a by now
familiar fashion, type IIA worldsheet instanton effects,
which fall off like $e^{-A_{IIA}/\alpha'}$, map to
field theory instanton effects on the heterotic side,
which fall off like $e^{-1/\lambda_{h,4}^2}$.  With nontrivial
$\pi_1$ on the type IIA side, there can be additional effects
which fall off like $e^{-R_{IIA}/\sqrt{\alpha'}}$, where
$R_{IIA}$ is the length of a nontrivial 1-cycle on C.  These
map to stringy instantons $e^{-h(M)/\lambda_{h,4}}$ on the heterotic
side, for an appropriate function of moduli $h(M)$.  

A more general origin for stringy instanton effects
comes from considering the heterotic/type I duality.
In ten dimensions, the map between the
couplings $\lambda_{h,10},\lambda_{I,10}$ and between the metrics
$g_{MN}^h,g_{MN}^I$ is simple:
$\lambda_h\leftrightarrow\lambda_{I,10}^{-1}$ and
$g_{MN}^h\leftrightarrow \lambda_h g_{MN}^I$ \eddyn.
Upon compactification, this indicates that worldsheet
instantons on one side map to effects that fall
off like $e^{-{A\over{\alpha'\lambda}}}$ on the other side, where
$A$ is the area of the 2-cycle about which the worldsheet wraps.
Such worldsheet instanton effects are ubiquitous
upon compactification.  Considering worldline instantons
in this context leads at least naively to even stronger
non-perturbative effects.  This will be discussed in 
\S2.4; the interpretation is not clear in terms of
large-order behavior of perturbation theory.

In both cases, the heterotic stringy non-perturbative
effects for appropriate quantities
can be computed in perturbation theory on
the dual side.  For compactifications with extended
supersymmetry ($N\ge 2$ in the $4d$ sense),
the effects can only appear in higher-derivative
(non-holomorphic) terms in the effective Lagrangian.
In the more generic case of $4d$ $N=1$ supersymmetry,
the effects can be seen in the K\"ahler potential.
There is in fact a restricted set of quantities
for which the non-perturbative effects can be
computed reliably, as will be discussed in \S2.  

One hope for the new understanding of non-perturbative
effects is that they can help with various longstanding puzzles
concerning the connection to observed low-energy physics.
In particular, duality between worldsheet instantons
and spacetime instantons has proved useful in computing
quantities (i.e. the spacetime superpotential) relevant
for lifting flat directions in moduli space and breaking
supersymmetry \refs{\gaugecond,\edsup,\kss,\edsupII,\ks,\mayr,\katzvafa}.
The known examples of type IIA/heterotic dual pairs include
a set \gaugecond\ which undergo gaugino condensation in a hidden
sector, potentially breaking supersymmetry.  These examples
are rather special, in that they arise as orientifolds of
known N=2 dual pairs, as in \vw.  However, as will be reviewed below,
the examples in \refs{\vw,\gaugecond}\
have the feature that the type IIA side has nontrivial
$\pi_1$, so the stringy instantons will appear.  
As discussed in \banksdine, it is difficult
to make this scenario of supersymmetry breaking work, but
the situation is improved given the possibility
for stringy instanton effects in the K\"ahler potential
(for discussion of some models see for example \gaillard\
and referenced therein).
In the type I/heterotic case, 
a nontrivial sector of the stringy instanton effects can 
be computed by making use of the inheritance in the
type I theory of worldsheet instanton sums computed
in the type IIB theory on the same space.

The paper is organized as follows.
In \S2, we explain the origin of the stringy instanton
effects from duality in the two cases, discuss
the class of amplitudes for which the analysis applies,
and raise a puzzle that appears in the heterotic/type I
context.  In \S3 we
discuss a few illustrative applications.

\newsec{Duality and Stringy Non-Perturbative Effects}

\subsec{Type I Worldsheet Instantons and $e^{-f(M)/\lambda_h}$ Effects}

Consider the type I string and the heterotic string compactified
on a D-dimensional manifold $M$.  As discussed in \eddyn, the
$10d$ metrics $g_{MN}^I$ and $g_{MN}^h$ and couplings
$\lambda_I$ and $\lambda_h$ are related by
\eqn\Ihet{g_{MN}^h=\lambda_hg_{MN}^I ~~~ \lambda_h={1\over\lambda_I}.}
In $10-D$ dimensions,  each string theory has worldsheet
instanton effects coming from the sector in the
$2d$ path integral where the string worldsheet
wraps around non-trivial 2-cycles.  In the heterotic string, on
a 2-cycle of area $A_h$, these fall
off like $e^{-A_h/\alpha'}$.  This maps to effects in
the type I string which fall off like 
$e^{-A_I/\alpha^\prime\lambda_I}$.  The interpretation is simple (and more
or less tautologous):  the heterotic string is a Dirichlet
1-brane in the type I string, and the heterotic worldsheet
instanton effects come from wrapping the D-1-brane worldsheet,
with tension $1/\lambda_I$, around the cycle of area $A_I$,
as in \bbs.  

In $4d$ (i.e. with $D=6$), there are field-theoretic
instanton effects in the heterotic string which fall off
like $e^{-1/\lambda_{h,4}^2}$, where $\lambda_{h,4}$ is
the $4d$ heterotic string coupling.  These map to
effects on the type I side which fall off like
$e^{-{V_I\over{(\alpha^\prime)^3\lambda_I}}}$,
where $V_I$ is the volume of $M$ measured in the type I metric.
This also has the simple interpretation of arising from
a wrapped Dirichlet 5-brane, in accord with the identification
of this object with the gauge instanton 5-brane in the
heterotic theory \sminst.

We can now go the other way.  On the type I side, there
are worldsheet instanton effects which arise from nontrivial
maps of the worldsheet into the target space.  For
2-cycles with area $A_I$, we find effects which fall off like
\eqn\Iws{\delta{\cal L}\sim e^{-{A_I\over{\alpha^\prime}}}.}
In terms of heterotic variables, this is
\eqn\Iwshet{\delta{\cal L}\sim e^{-{A_h\over{\lambda_h\alpha^\prime}}}.}
So each type I worldsheet instanton effect is a stringy
heterotic instanton effect.  Notice that we obtain
such effects without a stable solitonic type I string
from the heterotic point of view.  

In four-dimensional terms, the effects \Iwshet\ cannot appear in the
superpotential (or any holomorphic quantity).  This
agrees with the type I side due to the special
properties of the Ramond-Ramond anti-symmetric tensor
field $B_{MN}^I$.  This field cannot appear in non-derivative
couplings, so type I worldsheet instantons \Iws\ cannot 
correct the superpotential.  They can, however, affect
non-holomorphic quantities such as the K\"ahler potential
(and they do as we will review in \S3).  
We will argue that there is a set of quantities for which
the dual description provides a reliable computation
of the nonperturbative effects in \S2.3.

\subsec{Type IIA Worldline Instantons and Stringy Heterotic Instantons}

Consider a dual string pair with the property that
the inverse string coupling on one side maps to a radius on the
other side.  For simplicity let us consider
situations where the compactification manifold
preserves at least one covariantly constant spinor.  Then 
in terms of $4d$ $N=1$ superfields, we have
\eqn\couprad{S={1\over \lambda_4^2}+i{\theta}
\leftrightarrow T}
where $\lambda_4$ is the $4d$ string coupling, 
$\theta$ is the model-independent axion, and
$T$ is a chiral superfield made from a K\"ahler
modulus of the compactification.  As will be reviewed in
the next section, this is typical of four-dimensional
type IIA-Heterotic dual pairs with $N=1$, $N=2$, and
$N=4$ supersymmetry.  $T$ measures the
size of a holomorphic curve $C$ in the compactification
manifold, so that $Re(T)=\int_C J$, where $J$ is the K\"ahler form.
In such a situation, worldsheet instanton effects, which
fall off like 
\eqn\wsinst{\delta{\cal L}\sim e^{-T},}
map to spacetime instanton effects, which fall off like
\eqn\spinst{\delta{\cal L}\sim e^{-S}\sim e^{-1/\lambda_4^2}.}

Now consider the situation where the curve $C$ has nontrivial
fundamental group, $\pi_1\ne 0$.  That is, consider the case
where there is a nontrivial 1-cycle on $C$ of length
$L$.  Then there will
be worldsheet nonperturbative effects which fall off like
\eqn\wlinst{\delta{\cal L}\sim e^{-L}.}
In one channel this can be interpreted as arising
from a wrapped string, which is stable because of the
existence of a nontrivial cycle of length $L$  (not
because of any BPS condition).  
Let $L=h(M)\sqrt{Re(T)}$, where $h(M)$ is a function
of moduli, independent of $T$.
The dual then has the anticipated stringy behavior
\eqn\strinst{\delta{\cal L}\sim e^{-h(M)/\lambda_4}.}
It is interesting that particle-like behavior
(worldline instantons) maps to stringy behavior
($e^{-h(M)/\lambda_4}$ effects).

Notice that these stringy instantons disappear in
the decompactified 10-dimensional theory for
fixed 10-dimensional coupling $\lambda_{10}$, since
\eqn\tencoup{{1\over\lambda_4}={\sqrt{V}\over\lambda_{10}},}
where $V$ is the volume of the compactification.  So
the effects \strinst\ vanish as $V\to\infty$.  

Let us discuss the effects \wlinst\ more explicitly.
In a first quantized point particle framework, there is
a ``worldline instanton'' contribution
\eqn\wlII{\delta_m{\cal L}\sim \int [dX(\gamma)]
exp({-m\int_{\gamma}ds})}
for each state of mass $m$.  
This $e^{-{Lm}}$ behavior (for a loop of length $L$) can
be extracted from the 
$2d$ sigma model path integral.
Consider for example the computation
of the one-loop partition function for strings propagating
on a circle of radius R \joepart\mcclain.  The partition function is
\eqn\part{Z=\int_{{\cal F}} 
{d^2\tau\over\tau_2^2} Tr(q^{L_0}\bar q^{\bar L_0}).}
Here $\tau$ is the modular parameter for the
worldsheet torus, ${\cal F}$ is the fundamantal domain of the torus
moduli space, $q=e^{2\pi i\tau}$, and $L_0,\bar L_0$ are the
left and right-moving Hamiltonians on the worldsheet.
With supersymmetry the vacuum amplitude vanishes upon
summing over the spin structures.  Inserting vertex operators
into the correlation function will in general prevent
this cancellation, so let us concentrate on one sector, say
the (R,R) sector.  Then the normal ordering constant vanishes and
\eqn\Lzero{L_0={p^2\over 2}+ {p_L^2\over 2}+N_L=
{p^2\over 2}+{{1\over 2}({k\over{2 R}}+nR)^2}+N_L}
\eqn\bLzero{\bar L_0={p^2\over 2}+{ p_R^2\over 2}+N_R=
{p^2\over 2}+{{1\over 2}({k\over{2 R}}-nR)^2}+N_R.}
Here $N_L, N_R$ are the oscillator contributions, $p$ is
the spacetime momentum, and
$k,n$ are the momentum and winding numbers on the circle.
After trading the sum over $n$ for extending the integral
over the whole strip $-1/2<\tau_1<1/2$, $\tau_2>0$, and 
Poisson resumming the momentum mode sum, we find terms of the form
\eqn\Zterm{\int_0^{\infty} {d\tau_2\over\tau_2^{6}}
e^{-(m^2\tau_2+{R^2\over\tau_2})}}
where $m^2$ is the oscillator number $N$ on the string worldsheet.
The saddle point for the $\tau_2$ integral is at
$\tau_2=R/m$, leading to an effect which falls off like
\eqn\circ{\delta{\cal L}\sim e^{-mR}.}
In the $4d$ context with the relation \couprad,
this leads to a sum of terms in the dual theory
\eqn\addterms{\delta{\cal L}\sim
\sum_N{e^{-{\sqrt{N}\over\lambda_4}}a_N},}
where $a_N$ are possibly moduli-dependent coefficients.

\subsec{Regime of Validity}

In our analysis so far, we have simply mapped the
couplings and radii from one member of a dual pair
to the other; we then identified worldsheet or worldline
instanton effects on one side to effects nonperturbative
in the coupling on the other side.  In order for these
computations to be valid, we need (a subset of) each theory to be
weakly coupled in an appropriate sense, 
which we will now identify.

We have a compactification down to $d$ dimensions on both sides.
The $d$-dimensional couplings $\lambda_d$ and $\lambda^\prime_d$
can both be chosen to be weak in appropriate circumstances
\eddyn.  But the ten-dimensional couplings
$\lambda$ and $\lambda^\prime$ cannot both be weak.  
In the low-energy $d$-dimensional field theory, there are
loop effects which are suppressed by powers of 
$\lambda_d=\lambda/\sqrt{V}$ (here $V$ is the volume
of the compactification manifold).  There are in
general additional loop effects which occur with
powers of the ten-dimensional coupling $\lambda$
and no volume suppression.  These
are the loop effects which survive in the ten-dimensional limit.
At strong $10d$ coupling, these effects are large and cannot
be computed on both sides in a controlled way.  
So in order to compute reliably, we must restrict to
quantities which do not receive string loop corrections
in the decompactified $10d$ limit.  The set of such quantities
includes not only holomorphic objects in the $d$-dimensional
theory, but also terms in the effective action which
receive loop corrections only due to the supersymmetry breaking
provided by the compactification.  For example, the
kinetic terms for the massless fields 
coming from the $10d$ gravity multiplet in $4d$ $N=1$ compactifications
are in this class.  They are not protected from loop corrections
by holomorphy in $4d$, but in the $10d$ limit they reduce
to kinetic terms for the gravity multiplet, which are fixed
by $10d$ $N=1$ supergravity.  On the other hand, an arbitrary
scattering amplitude is not protected from loop effects
in any dimension.  So although the full theory is not under control
from both sides of the dual pair, amplitudes which are
protected in the $10d$ limit can be computed reliably.

\subsec{Type I/Heterotic Duality Revisited:  A Puzzle}

Considering worldline instanton effects of
the type discussed in \S2.3 in the type I/heterotic context suggests
even stronger instanton effects on both sides.
For this duality, we have so far concentrated on effects which fall
off like $e^{-{{\rm Area}\over\alpha^\prime}}$ on one side or the other.
Consider the special case that the compactification
manifold $M$ has nontrivial fundamental
group.  Then as in \S2.2, at the perturbative level on each side
there are effects which fall off like 
$e^{-{R\over{\sqrt{\alpha^\prime}}}}$, where
$R$ is the radius of the nontrivial 1-cycle.  These
effects map to effects on the other side which  
fall off like 
$e^{-{R^\prime\over{\sqrt{\alpha^\prime}(\lambda^\prime)^{1\over 2}}}}$.  
These are stronger at weak coupling than the effects
predicted in \steve.
Perhaps this is related to the fact that there is
a factor of $R$ in the exponent, so that the effect
is suppressed by large radius as well as by weak coupling.
It would be very interesting to 
understand the interpretation of these effects
(if they do not always cancel somehow in the reliably computable amplitudes)
in terms of the large-order behavior of perturbation theory.

\newsec{Some Applications}

\subsec{heterotic/type I}

The ten-dimensional heterotic/type I duality holds upon compactification
on a smooth D-dimensional manifold
$M$.  One can consider the effects of
\S2.1 on anything from the highly symmetric $T^D$ to a 
generic Calabi-Yau (or even a manifold of exceptional
holonomy).  Of particular interest are Calabi-Yau three-folds,
for which $4d$ $N=1$ supersymmetry is preserved perturbatively
in $\alpha^\prime$ and $\lambda$.  
Worldsheet instanton effects coming from sphere
amplitudes in the type I theory are
inherited from those in the type IIB string theory.
These can be computed exactly for some terms
in the effective action using mirror symmetry,
as in \cdgp. 
One such term is the K\"ahler potential for
moduli fields of the manifold.  (In the type I
context, the gauge bundle moduli as well as
the charged fields are open strings and do not
couple on the sphere.)    

On the quintic hypersurface in ${\bf CP}^4$ for example,
the worldsheet instanton sum has been computed
in terms of the single K\"ahler modulus $T_I$ \cdgp.
In terms of the holomorphic prepotential ${F}$ in
the type IIB compactification on the Calabi-Yau, the
K\"ahler potential for $T_I$ is (in ``special coordinates'')
\specgeom
\eqn\kp{K(T_I,\bar T_I)=-\log\biggl[2(F(T_I)+F(\bar T_I))
-(T_I-\bar T_I)(F_{T_I}-\bar F_{\bar T_I})\biggr]}
where $F_{T_I}={{\partial F}\over{\partial T_I}}$. 
From \cdgp\ we have
\eqn\yuk{{\partial^3{F}\over{\partial T_I^3}}
\sim 5+\sum_{k=1}^\infty{n_kk^3e^{-2\pi kT_I}
\over{1-e^{-2\pi kT_I}}}}
where $n_k$ is the number of degree $k$ holomorphic curves.
Integrating this and plugging into \kp\ gives the
K\"ahler potential.  Substituting $Re(T_h)/\lambda_h$ for
$T_I$ into the instanton contributions gives the
type I sphere contribution to the
stringy instanton sum in the K\"ahler potential on the heterotic side.
This correspondence
of course applies to all smooth Calabi-Yau compactifications.
It would be interesting to construct models with dynamical supersymmetry
breaking and check whether this contribution to the K\"ahler
potential actually satisfies the conditions for stabilizing
the dilaton discussed in \refs{\banksdine,\gaillard}.  It is intriguing
that this contribution to the K\"ahler potential takes the $N=2$ form
in this $N=1$ theory.

\subsec{heterotic/type IIA}

The simplest example of the type of duality discussed in
\S2.2 has $N=4$ supersymmetry in $4d$.  One side of the pair 
is the type IIA theory
on $K3\times T^2$, and the other is the 
heterotic theory on $T^6\sim T^4\times T^2$.
As explained in \eddyn, the dilaton $S$ gets exchanged
with the $T^2$ K\"ahler modulus $T$ under the duality as
in \couprad.  Furthermore the $T^2$ has nontrivial
$\pi_1$.  In this case, although terms like 
\Zterm -\addterms\ appear, they cancel by holomorphy in some
one-loop amplitudes (see for example the computation
of threshold corrections from the untwisted sector of
orbifolds in \dkl, and an analogous computation
in $N=4$ theories in \harvmoore).  
It might be interesting to look in this
$N=4$ context for processes which 
satisfy the reliability criteria discussed in \S2.3,
but which are not protected by holomorphy in $4d$.

In any case, let us proceed to the more generic models with less supersymmetry.
The $4d$ $N=2$ supersymmetric dual string pairs of
\refs{\kv,\fhsv}\ also satisfy \couprad.  These examples, as
well as others discovered subsequently, consist
of the type IIA string theory compactified on a $K3$ fibration
on one side and the heterotic theory on $K3\times T^2$ on the other.
The $K3\times T^2$ on the heterotic side can itself be
thought of as a $T^4$ fibration, so that the whole dual
pair is a fibration of the six dimensional duality between
type IIA on $K3$ and the heterotic theory on $T^4$
\refs{\klm,\vw,\asplouis}. 
The base of this fibration is a ${\bf P}^1$ whose area
is given by the real part of 
the K\"ahler modulus T.  All of these models
have $S\leftrightarrow T$, where $S$ is the dilaton 
superfield.  However, since
the base ${\bf P}^1$ is simply connected, the effects discussed
above will not occur.

For $4d$ $N=1$ supersymmetric models obtained from the
$N=2$ dual pairs by orientifolding \refs{\vw,\gaugecond}, nontrivial
$\pi_1$ is generated on the base.  The orbifold group
acts on the base ${\bf P}^1\sim S^2$ by identifying
antipodal points on the sphere.  This leaves 
${\bf RP}^2$, with 
$\pi_1={\bf Z}_2$.  Hence for these examples, both of
the conditions in \S2 are satisfied, and we expect
nontrivial stringy instanton effects on the heterotic side.

\smallskip
{\bf Acknowledgements:}
I would like to thank T. Banks and S. Shenker for very
helpful discussions.  I am also grateful to P. Aspinwall,
S. Kachru, and J. Maldacena for useful comments.  This work was
supported in part by DOE grant DE-FG02-96ER40559.

\listrefs
\end